\documentclass[aps,prl,twocolumn,amsmath,amssymb,notitlepage,noeprint,nofootinbib,superscriptaddress]{revtex4-1}
\pdfoutput=1
\usepackage{braket}
\usepackage{xcolor}
\usepackage{verbatim}
\usepackage{graphicx}
\usepackage{esint}
\usepackage{epstopdf}
\usepackage{float}
\usepackage{bm}
\definecolor{darkblue}{rgb}{0,0,0.6}
\usepackage[colorlinks,linkcolor=darkblue,citecolor=darkblue,urlcolor=darkblue]{hyperref}
\newcommand{\beq}{\begin{equation}} \newcommand{\eeq}{\end{equation}}
\newcommand{\argc}[1]{\left[#1\right]}

\let \a=\alpha
\newcommand{\ta}{\tau_{\alpha}}

\begin{document}

\title{Microscopic origin of excess wings in relaxation spectra of supercooled liquids}

\author{Benjamin Guiselin*}

\affiliation{Laboratoire Charles Coulomb (L2C), Universit\'e de Montpellier, CNRS, 34095 Montpellier, France}

\author{Camille Scalliet*}

\affiliation{Department of Applied Mathematics and Theoretical Physics, University of Cambridge, Wilberforce Road, Cambridge CB3 0WA, United Kingdom}

\author{Ludovic Berthier}

\affiliation{Laboratoire Charles Coulomb (L2C), Universit\'e de Montpellier, CNRS, 34095 Montpellier, France}

\affiliation{Department of Chemistry, University of Cambridge, Lensfield Road, Cambridge CB2 1EW, United Kingdom}

\let\thefootnote\relax\footnotetext{*Equal contributions.}

\date{\today}

\begin{abstract}
Glass formation is encountered in diverse materials. Experiments have revealed that dynamic relaxation spectra of supercooled liquids generically become asymmetric near the glass transition temperature, $T_g$, where an extended power law emerges at high frequencies. The microscopic origin of this ``wing'' remains unknown, and was so far inaccessible to simulations. Here, we develop a novel computational approach and study the equilibrium dynamics of model supercooled liquids near $T_g$. We demonstrate the emergence of a power law wing in numerical spectra, which originates from relaxation at rare, localised regions over broadly-distributed timescales. We rationalise the asymmetric shape of relaxation spectra by constructing an empirical model associating heterogeneous activated dynamics with dynamic facilitation, which are the two minimal physical ingredients revealed by our simulations. Our work offers a glimpse of the molecular motion responsible for glass formation at relevant experimental conditions.
\end{abstract}

\maketitle

The formation of amorphous solids results from the rapid growth of the structural relaxation time $\tau_\alpha$ of the supercooled liquid~\cite{berthier2011theoretical}. Molecular motion occurs on a timescale of about $10^{-10}$~s at the onset temperature of glassy behaviour but takes about $100$~s at the experimental glass transition temperature $T_g$~\cite{schmidtke2012boiling}. Over the last decades, dielectric, mechanical and light scattering experiments kept developing to probe molecular motion over a broader frequency range with increased accuracy~\cite{lunkenheimer2000glassy,rossler,schmidtke2013reorientational,gainaru2009evolution,flamig2020nmr,Hecksher}. This progress reveals that the temperature evolution of $\tau_\alpha$ is just the tip of the iceberg, as relaxation spectra $\chi''(\omega)$ measured near $T_g$ exhibit relaxation processes taking place over an extremely large frequency window~\cite{schneider2000excess,lunkenheimer2002excess,nagel_scaling,leheny1998dielectric}. The overall shift of relaxation spectra is accompanied by an equivalent broadening of about 12 decades, which is the other side of the same coin. A microscopic explanation of these slow dynamics is at the heart of glass transition research~\cite{berthier2011theoretical}. 

High-temperature spectra reflect near exponential relaxation in the picosecond range, but low-$T$ spectra broaden into a two-step process with a stretched exponential relaxation at low frequency $\omega \approx 1/\ta$ and a microscopic peak remaining at the picosecond timescale. In 1990, Nagel and coworkers~\cite{menon1992wide,nagel_scaling,nagel_scaling2,leheny1998dielectric} showed that for a number of molecular liquids the structural relaxation peak extends much further at high frequencies $\omega \tau_\alpha \gg 1$ and transforms into a power law, $\chi''(\omega) \sim \omega^{-\sigma}$, with a small exponent $\sigma(T)\in\argc{0.2,0.4}$ decreasing with temperature~\cite{nagel_scaling2}. Using logarithmic scales, this resembles a ``wing'' in ``excess'' of the $\alpha$-peak. At $T_g$, the wing extends over the mHz-MHz range with an amplitude about 100 times smaller than the $\alpha$-peak. A universal scaling comprising the excess wing was proposed~\cite{nagel_scaling2}, which can be altered by additional microscopic processes~\cite{wu1991relaxation,ngai_classification}. While this universality is debated~\cite{schneider2000excess,lunkenheimer2002excess}, the presence of an excess contribution often taking the form of a wing is not~\cite{blochowicz2003susceptibility,gainaru2009evolution}.

Elucidating the nature of molecular motions responsible for the small signal in these excess wings appears daunting. Yet, experiments managed to characterise its heterogeneous nature~\cite{bauer2013nonlinear,duvvuri2003dielectric} and aging properties~\cite{lunkenheimer2005glassy}. So far, computer simulations were unable to access the required range of equilibration temperatures and timescales to even address the question. Physical interpretations and empirical models have been proposed to explain the shape of relaxation spectra. Some of them couple slow translational motion to an ``additional'' degree of freedom ({\it e.g.}, rotational)~\cite{diezemann1999slow,domschke2011glassy}. Others invoke spatially heterogeneous dynamics to construct a broad distribution of timescales of static~\cite{sethna1991scaling, stevenson2010universal, viot2000heterogeneous, chamberlin1999mesoscopic, dyre2000universality} or kinetic~\cite{berthier2005numerical} origin. The winged asymmetric shape then requires specific physics, such as geometric frustration~\cite{viot2000heterogeneous}, lengthscale-dependent dynamics~\cite{chamberlin1999mesoscopic}, or dynamic facilitation~\cite{berthier2005numerical}. With specific choices, these approaches yield relaxation spectra comprising excess wings, but direct microscopic investigations testing the underlying hypotheses are still lacking.

Here, we show that computer simulations can now directly observe excess wings and assess their microscopic origin. We take advantage of the recent swap Monte Carlo algorithm~\cite{swap} to efficiently produce equilibrated configurations of a supercooled liquid with $\tau_\alpha \approx 100$~s. We observe their physical relaxation dynamics over ten decades in time, up to $20$~ms. We are thus able to probe for the first time the temperature and time regimes where excess wings are observed in experiments. We report the emergence of a power law (a wing) in numerical spectra with the same characteristics as in experiments. We demonstrate that it is caused by a sparse population of localised regions, whose relaxation times are power law distributed. These relaxed regions then coarsen by dynamic facilitation. We construct an empirical model to illustrate how heterogeneous dynamics and dynamic facilitation generically lead to asymmetric, winged relaxation spectra.

\begin{figure}[t]
\includegraphics[width=\columnwidth]{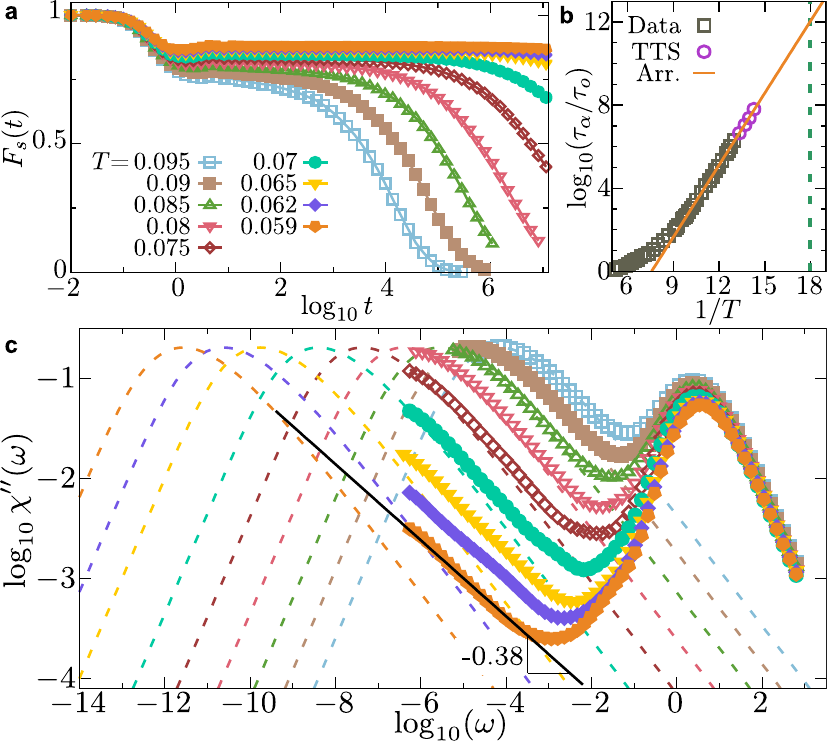}
\caption{ {\bf Emergence of excess wings in a three-dimensional glass-former near the glass transition temperature.} (a) Self-intermediate scattering function $F_s(t)$ at various temperatures. (b) Relaxation time $\ta$ rescaled by its value $\tau_o$ at the onset temperature. Symbols are directly measured data (squares), or obtained using time temperature superposition (TTS, circles). A conservative Arrhenius extrapolation locates $T_g = 0.056$, where $\ta(T_g)/\tau_o=10^{12}$ (dashed line). (c) Relaxation spectra for the same temperatures as in panel (a). The dashed lines represent the estimated $\a$-peaks. Close to $T_g$, the spectra lie above the $\alpha-$peak and display a power law signal with an exponent $\sigma\approx 0.38$ (full line), in quantitative agreement with excess wings observed experimentally.}
\label{fig:1}
\end{figure}

\begin{figure}[t]
\includegraphics[width=\columnwidth]{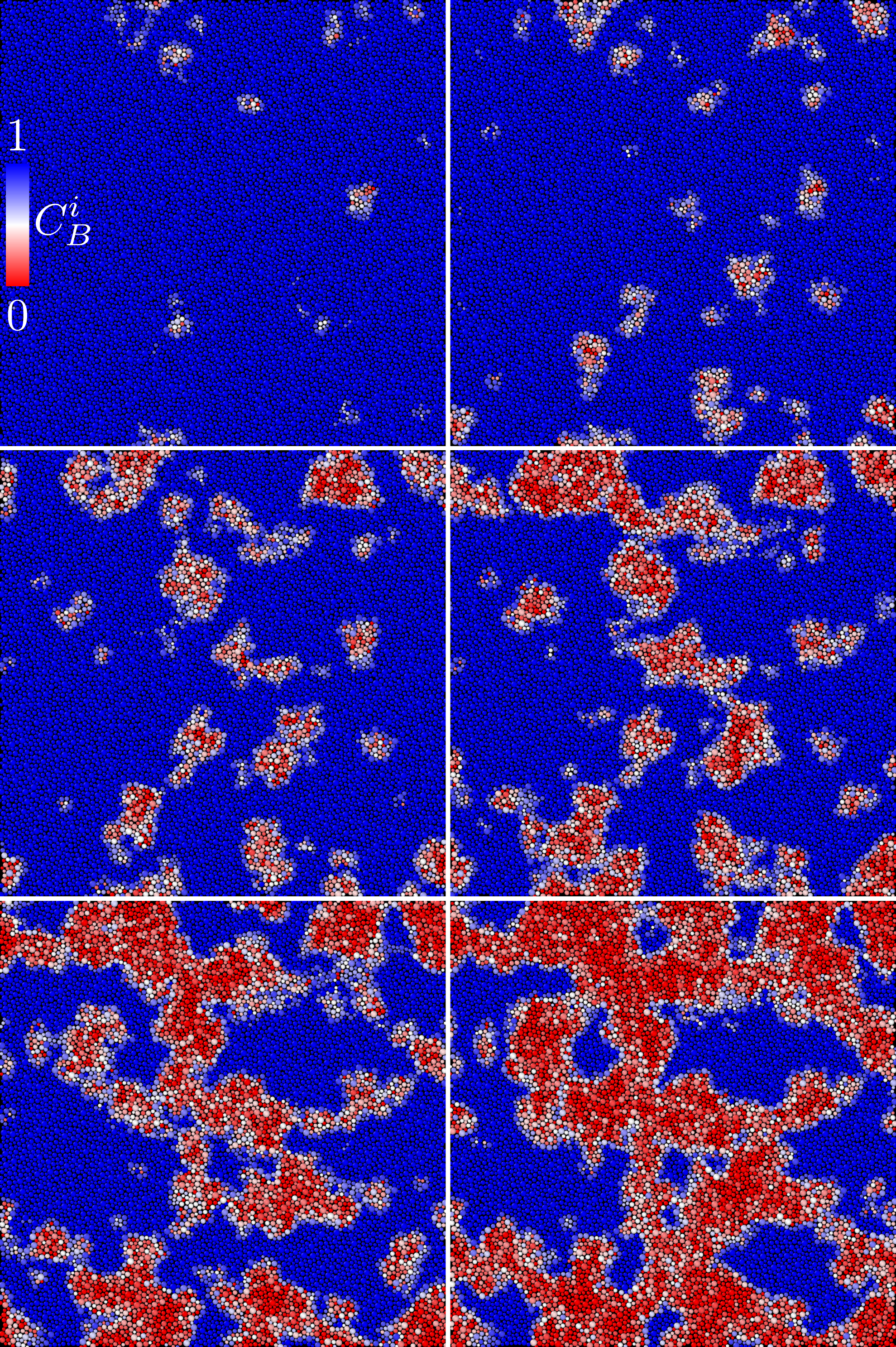}
\caption{{\bf Visualisation of spatially heterogeneous and facilitated dynamics.} Relaxation in the $2d$ system at $T_{2d}=0.09$ with $\ta/\tau_o =10^8$. Frames are logarithmically spaced between $t = 2 \times 10^{-3} \tau_{\a}$ (top left) and $t=0.6 \ta$ (bottom right) from left to right and top to bottom. Particles are coloured according to $C_B^i(t)$ from blue [immobile, $C_B^i(t)=1$] to red [relaxed, $C_B^i(t)=0$]. The linear size of the simulation box is 100.}
\label{fig:2}
\end{figure}

\begin{figure}[t]
\includegraphics[width=\columnwidth]{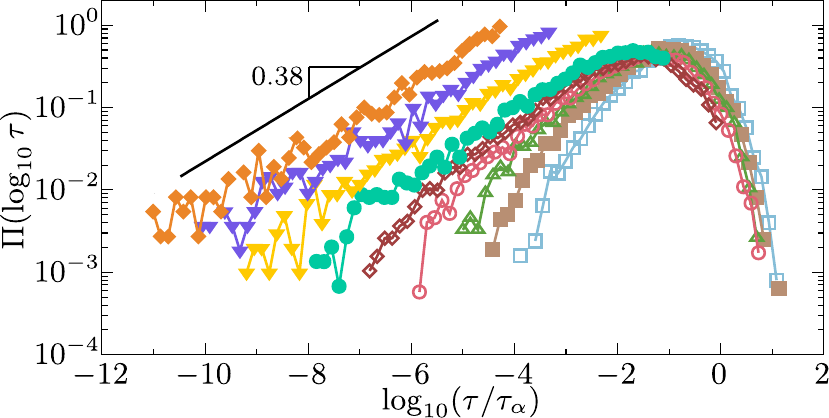}
\caption{{\bf Microscopic origin of excess wings.} Waiting-time distribution of newly relaxing clusters in $3d$ from $T_{\rm mct}$ (right) to $T_g$ (left), with the same color code as Fig.~\ref{fig:1}(a). Approaching $T_g$, the distributions develop a power law tail at $\tau \ll \tau_{\alpha}$, with an exponent $0.38$ that directly accounts for the excess wings in the spectra of Fig.~\ref{fig:1}(c).}
\label{fig:3}
\end{figure}

We study size-polydisperse mixtures of $N$ soft repulsive spheres in two and three dimensions, as described in the \hyperref[sec:methods]{Methods} section. These models are representative computational glass-formers~\cite{berthier2019zero,berthier2017configurational}. We use the swap Monte Carlo algorithm designed in Ref.~\cite{lammpsswap} to generate $n_s\in \argc{200,450}$ independent equilibrium configurations at temperatures $T$ down to the extrapolated experimental glass transition temperature $T_g$. Each equilibrium configuration is then taken as the initial condition of a multi-CPU molecular dynamics (MD) simulation (without swap). The $n_s$ independent simulations run for up to a simulation time $t_{\textrm{max}} = 1.5\times 10^{7}$ in $3d$ (one week on 2 CPUs for $N=1200$). We push a few $2d$ simulations to unprecedentedly long times, up to $t_{\textrm{max},2d}  = 6\times 10^{8}$, representing a computational time of several months. By using the relaxation time at the onset of glassy dynamics to relate numerical and experimental timescales, our longest simulations translate into a physical time of about $20$~ms for systems having an equilibrium relaxation time $\tau_\alpha \approx 10^2$~s. This strategy is key to observe excess wings, which would otherwise be buried underneath the structural relaxation in conventional approaches~\cite{yu2017structural}. The $2d$ and $3d$ models behave similarly, so we present quantitative results for the $3d$ model ($N=1200$) in Figs.~\ref{fig:1},~\ref{fig:3} and illustrate the relaxation process in Fig.~\ref{fig:2} with $2d$ snapshots  ($N = 10000$), which are easier to visualize. Quantitative results for the $2d$ model are provided in the Supplementary Information (SI).

We investigate the spatio-temporal evolution of the relaxation dynamics using averaged and particle-resolved dynamic observables. In $3d$, we measure the self-intermediate scattering function
$F_s(t)$, averaged over the $n_s$ independent runs. We define the relaxation time $\ta$ by $F_s(\ta)=e^{-1}$. In $2d$, collective long-ranged fluctuations affect the measurement of $F_s(t)$. We instead focus on observables which are blind to these fluctuations~\cite{illing2017mermin} and define $\tau_{\a}$ via the bond-orientational correlation function~\cite{flenner2015fundamental}. In both two and three dimensions, we investigate the relaxation process at the particle scale via the bond-breaking correlation $C_B^i(t)$ which quantifies the fraction of nearest neighbours lost by particle $i$ after time $t$. Starting from $C_B^i(t=0)=1$, it decreases as rearrangements take place close to particle $i$, and reaches zero when its local environment is completely renewed. Precise definitions of the correlation functions are provided in the \hyperref[sec:methods]{Methods} section.

To connect with experimental results obtained in the frequency domain, we compute a dynamic susceptibility $\chi''(\omega)$ from a distribution of relaxation times $G(\log \tau)$~\cite{blochowicz2003susceptibility,berthier2005numerical}
\begin{equation}
\chi''(\omega)=\int_{-\infty}^{\infty} G(\log \tau) \frac{\omega \tau }{1+(\omega \tau)^2}\mathrm{d}\log \tau~~,
\label{eq:chi}
\end{equation} 
where the distribution $G$ is related to the derivative of a time correlation function, $G(\log t) \approx -\mathrm{d}F_s(t)/\mathrm{d}\log t$ in $3d$. We use the bond-breaking correlation function instead of $F_s$ in $2d$. We discuss the numerical evaluation of $\chi''$ in the \hyperref[sec:methods]{Methods} section, whereas the discussion on the statistical noise and the comparison to direct Fourier transforms are in the Supplementary Information, Section I and Figure \ref{fig:2}.

We start by presenting equilibrium measurements of $F_s(t)$ in $3d$ in Fig.~\ref{fig:1}(a), concentrating on the unexplored low-$T$ regime below the mode-coupling crossover $T_{\rm mct} \approx 0.095$. The latter is determined by a power law fit of $\ta(T)$ in the range $\ta/\tau_o < 10^3$, where $\tau_o \approx 3$ is the value of $\ta$ at the onset temperature $T_o \approx 0.20$~\cite{swap}. At all temperatures, the correlations display a fast initial decay near $t \approx \tau_o$, due to fast dynamical processes. At larger times, we observe a much slower decay to zero. As $T$ decreases, the relaxation time grows and eventually exits the numerically accessible time window. At the lowest investigated temperatures near $T_g$, correlations appear almost constant over more than 7 decades in time, suggesting near-complete dynamic arrest. We recall that thanks to the swap algorithm, all measurements reflect genuine equilibrium dynamics, even when $\ta$ is larger than the simulated time by many orders of magnitude.

Our strategy allows us to directly observe the $\alpha$-relaxation when $\ta < t_{\textrm{max}}$, equivalently $\ta/\tau_o \lesssim 5 \times 10^6$ down to $T=0.0755$, see Figs.~\ref{fig:1}(a,b). In this regime, the relaxation is well-described by a stretched exponential $F_0 e^{-(t/\ta)^\beta}$ with an almost constant stretching exponent $\beta \approx 0.56$, the amplitude $F_0$ modestly changing with temperature. We use this time temperature superposition (TTS) property to estimate $\ta$ for $0.07 \leq T \leq 0.0755$, where the decorrelation of $F_s(t)$ is sufficient~\cite{howtomeasure}, and obtain $\ta$ over roughly 2 additional decades, see Fig.~\ref{fig:1}(b). We finally use an Arrhenius law to extrapolate $\ta$ over 4 more decades to get a safe lower bound for the experimental glass temperature $T_g \approx 0.056$, defined by $\tau_\a(T_g)/\tau_o=10^{12}$~\cite{swap}, see the \hyperref[sec:methods]{Methods} section for details.

The corresponding relaxation spectra are shown in Fig.~\ref{fig:1}(c) for the $3d$ model. They all display a peak at high frequency $\omega \approx 1/\tau_o$, corresponding to the short-time decay of $F_s(t)$. A low-frequency peak near $\omega \approx 1/\tau_\a$ is also visible. As $T$ decreases, this $\alpha$-peak shifts to lower frequencies and eventually exits the accessible frequency window. When the $\alpha$-peak is not directly measured, we extrapolate its shape by inserting the above stretched exponential form for $F_s(t)$ into Eq.~(\ref{eq:chi}). We use $\beta = 0.56$, $\ta$ given by the Arrhenius extrapolation, and a constant $F_0$. The tiny temperature dependence of $F_0$ is immaterial on the logarithmic scale of Fig.~\ref{fig:1}(c). The resulting $\alpha$-peaks are shown in Fig.~\ref{fig:1}(c) with dashed lines that smoothly merge into the measured data at the highest temperatures, validating our procedure. 
 
As $T$ decreases, the measured susceptibility and the $\alpha$-peak deviate increasingly from one another, the data being systematically in excess of the $\alpha$-peak. Since the Arrhenius extrapolation underestimates $\ta$, this excess is (at worst) slightly underestimated and cannot be accounted by a vertical shift which would require unphysical values of $F_0$ and $\beta$. At the lowest $T$, where the $\alpha$-peak no longer interferes with the measurements, the spectra are well described by a power law $\chi^{\prime\prime}(\omega)\sim \omega^{-\sigma}$ at low frequencies, with an exponent $\sigma \approx 0.38$ slightly decreasing with $T$, and an amplitude about 100 times smaller than the $\alpha$-peak. The relaxation spectra of the $2d$ model in Supplementary Figure 3 exhibit similar features with an exponent $\sigma_{2d} \approx 0.45$, which is quite close to the one found in $3d$. In our simulations, the measured spectra do not exhibit a secondary peak separated from the $\alpha$-relaxation, and cannot be interpreted using an additive $\beta$-process~\cite{yu2017structural}. Therefore, close to $T_g$, the numerical spectra follow a power law over a similar frequency range, with a similar exponent and a similar amplitude as the excess wings obtained experimentally, suggesting that simulated glass-formers display excess wings resembling observations in molecular liquids. 

We take advantage of the atomistic resolution offered by simulations to explore the microscopic origin of excess wings and provide a physical interpretation of the spectral shapes.We illustrate the relaxation dynamics with $2d$ snapshots, which are easier to render and interpret. We confirm that the same mechanisms are observed in $3d$. In Fig.~\ref{fig:2} we show $2d$ snapshots illustrating how structural relaxation proceeds at a temperature $T_{2d} = 0.09$ (we estimate $T_{g,2d} \approx 0.07$) for which $\ta/\tau_o \approx 10^8$, corresponding to around 10~ms in physical time. This temperature is the lowest for which the $\alpha$-relaxation can be observed in the numerical window, and is considerably lower than the mode-coupling crossover near $T_{{\rm mct},2d}\approx 0.12$. Images are shown at logarithmically-spaced times $t$ in the range $t /\ta\in \argc{10^{-3},1}$. Particles are coloured according to $C_B^i(t)$: red particles have relaxed, blue ones have not. We present in Fig.~S3 the relaxation spectrum measured at this temperature.

For $t \ll \ta$, relaxation starts at a sparse population of localised regions which emerge independently throughout the sample over broadly distributed times. This conclusion holds over a large range of temperatures down to $T_g$ in both $d=2,3$. As time increases, newly relaxed regions continue to appear, but a second mechanism becomes apparent in Fig.~\ref{fig:2} as regions that have relaxed in one frame typically appear larger in the next. This growth of relaxed regions in Fig.~\ref{fig:2} is the signature of dynamic facilitation~\cite{chandler2010dynamics}. More precisely, we observe that from one frame to the next, relaxation events keep accumulating at similar locations, which results in mobile particles undergoing multiple relaxations and mobility propagating to nearby particles. Also, the slowest regions are typically ``invaded'' at $t \gg \ta$ from their faster boundaries. Dynamic facilitation has been identified before at high temperatures above the mode-coupling crossover~\cite{chandler2010dynamics, keys2011excitations, vogel2004spatially}. Our investigations show that it becomes a central physical mechanism for structural relaxation near $T_g$. 

We concentrate on the early times where power law spectra are observed. Visualisation suggests that clusters of relaxed particles appear at sparse locations. We now establish that these early relaxation events are responsible for the excess wing. To this end, we define mobile ($C_B^i < 0.55$) and immobile ($C_B^i\geq 0.55$) particles; the threshold value near 0.5 is determined requiring self-consistency with alternative mobility definitions based on displacements. We identify connected clusters of mobile particles by performing a nearest neighbour analysis (details in the \hyperref[sec:methods]{Methods} section), and investigate the statistical properties of relaxed clusters. In particular, we find that the excess wing regime at $t/\ta\ll 1$ is dominated by the appearance of new clusters, whereas the growth of existing clusters dominates at later times. We report in Fig.~\ref{fig:3} the distribution $\Pi(\tau)$ of waiting times $\tau$ for the appearance of new clusters in $3d$. For $T\leq 0.07$, we cannot measure the entire distribution, which is thus determined up to an uninteresting prefactor. The corresponding $2d$ results are shown in Supplementary Figure 4.

At the highest investigated temperature, near $T_{\rm mct}$, the distribution $\Pi(\log_{10} \tau)$ in Fig.~\ref{fig:3} is already very broad, with clusters appearing as early as $10^{-4}\ta$. The distribution peaks near $0.1\ta$, when dynamic facilitation starts to dominate, and has a cutoff around $10\ta$. As $T$ decreases below the mode-coupling crossover, a power law tail emerges at $\tau \ll \ta$. For $T \leq 0.07$, the power law extends over at least 6 decades, with a nearly constant exponent $\Pi(\log_{10}\tau) \sim \tau^{0.38}$ for the $3d$ model. The relaxation of localised clusters at early times is extremely broadly distributed, presumably stemming from an equally broad distribution of activation energies. 

Remarkably, if we plug the measured distribution of waiting times in Eq.~(\ref{eq:chi}), a power law $\Pi(\log_{10}\tau) \sim \tau^{0.38}$ directly translates into a power law $\chi''(\omega)\sim \omega^{-0.38}$ in the spectra, which is thus valid for $\omega \ta \gg 1$. The agreement with the data in Fig.~\ref{fig:1}(c) is therefore quantitative. A similar agreement is found in $2d$ with the exponent $\sigma_{2d}=0.45$, see Supplementary Figure 4. This analysis demonstrates that the high-frequency power law in $\chi''(\omega)$ stems from the relaxation of a sparse population of clusters characterised by a broad distribution of relaxation times.

\begin{figure}[t]
\includegraphics[width=0.85\columnwidth]{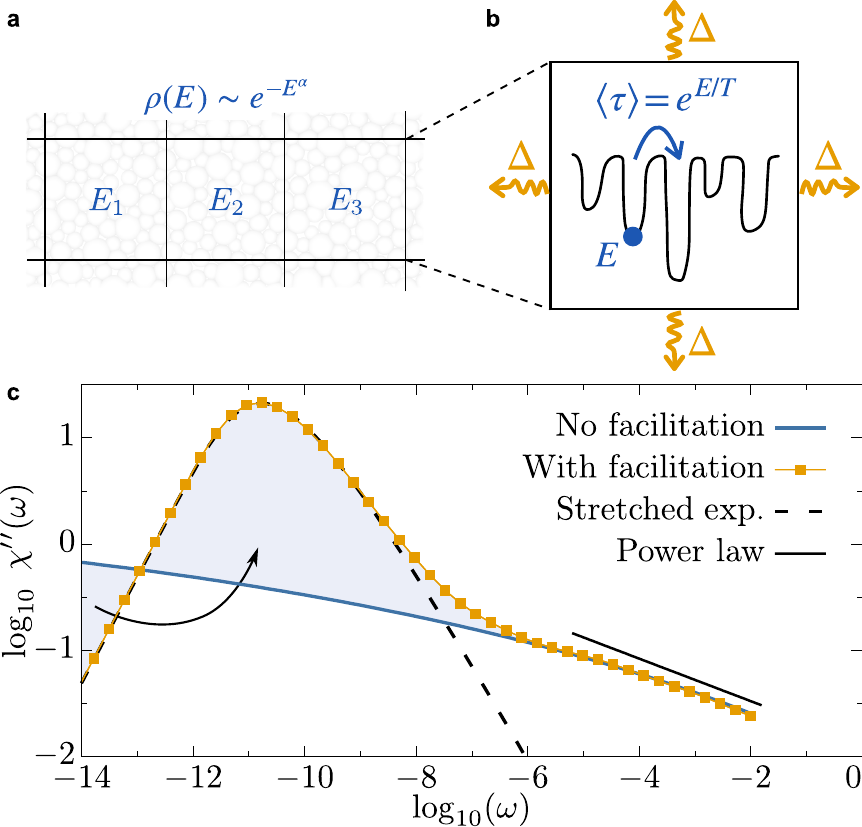}
\caption{{\bf Facilitated trap model generically predicts asymmetric winged relaxation spectra.} (a) The liquid is modeled as a collection of traps with energies $E$, distributed according to $\rho(E)$. (b) Relaxation is thermally activated and affects the energy of the other traps by a random amount, proportional to $\Delta$. (c) Relaxation spectra $\chi ''(\omega)$ in the absence ($\Delta = 0$) and presence ($\Delta=0.05$) of dynamic facilitation at $\a=1.1$ and $T=0.629$. Dynamic facilitation compresses (arrow) the low-frequency part of the underlying spectrum, giving rise to a sharper $\alpha$-peak well-fitted by the spectrum of a stretched exponential (dashed). The high-frequency part of the underlying spectrum, unaffected by facilitation, is well described by a power law $\omega^{-0.2}$ (line). The excess wing thus corresponds to the beginning of the relaxation process.}
\label{fig:4}
\end{figure}

This microscopic view of the power law wing alone does not explain why it appears in excess of the $\alpha$-peak observed at larger times when dynamic facilitation sets in. To explain this point, we construct an empirical model based on our numerical observations. We first imagine that the liquid can be decomposed into independent domains characterised by a local relaxation time, see Fig.~\ref{fig:4}(a). This heterogeneous viewpoint is mathematically captured by trap models~\cite{dyre1987master,bouchaud1992weak}. To introduce dynamic facilitation as the second key ingredient, we construct a facilitated trap model, assuming that a given local relaxation event may now affect the state of the other traps, see Fig.~\ref{fig:4}(b). To provide a qualitative, generic description of relaxation spectra, we analyse the simplest version of such a model and assume, in a mean-field spirit, that dynamic facilitation equally affects all traps. A more local version was designed in Refs.~\cite{rehwald2010coupled,rehwald2012coupled} for different purposes.

We consider $N$ traps with energy levels $E>0$ drawn from a distribution $\rho(E)$, and assume activated dynamics. The energy $E$ of a trap is renewed after a Poisson-distributed timescale of mean $\langle \tau(E) \rangle = e^{E/T}$. Since deep traps take much longer to relax than shallow ones, the system is dynamically heterogeneous. Following Ref.~\cite{arkhipov1994random}, we use $\rho(E)\propto e^{-E^\a}$, with $\a \in \argc{1,2}$ to smoothly interpolate between the much-studied Gaussian~\cite{dyre1987master,rehwald2010coupled} and exponential~\cite{bouchaud1992weak} distributions. Dynamics at temperature $T$ leads to the equilibrium energy distribution $P_{\text{eq}}(T,E) \propto \rho(E) e^{E/T}$. Whenever a trap relaxes, the energy of all other traps is shifted by a random amount uniformly distributed in the interval $\argc{- \frac{\Delta}{\sqrt{N}}, \frac{\Delta}{\sqrt{N}}}$, using a Metropolis filter to leave the equilibrium distribution $P_{\text{eq}}$ unchanged. This coupling between traps mimics dynamic facilitation~\cite{rehwald2010coupled}. The relaxation spectra $\chi''(\omega)$ is computed either analytically ($\Delta = 0$), or by simulating the facilitated model ($\Delta > 0$).

The model is specified by two parameters $(\a,\Delta)$, for which equilibrium dynamics can be studied at any temperature $T$. We have systematically investigated this parameter space, and find spectra with quantitative differences but generic features~\cite{doi:10.1063/5.0060408}. In Fig.~\ref{fig:4}(c), we select $(\a=1.1, \Delta = 0.05)$ at $T=0.629$ for aesthetic reasons, as this produces a spectrum qualitatively resembling experimental and numerical ones close to $T_g$. Fitting the $\alpha$-peak to the frequency representation of a stretched exponential reveals an excess wing at high frequencies. However, in the absence of dynamic facilitation ($\Delta=0$) one obtains the blue spectrum, with the same high-frequency behaviour, but which extends much further at low frequencies. Indeed, without facilitation each trap relaxes independently, and the equilibrium distribution $P_{\rm eq}$ determines the dynamic spectrum, which is broad and relatively symmetric. In the presence of facilitation, $\Delta > 0$, shallow traps still relax independently and are essentially unaffected. Crucially, deep traps now receive small kicks whenever a shallow trap relaxes, and their energies slowly diffuse towards the most probable value. This accelerates their relaxation, which eventually affects the tail of the relaxation time distribution. As a result, dynamic facilitation ``compresses'' the low-frequency part of the underlying spectrum (blue), as hinted in Ref.~\cite{xia2001microscopic}, and highlighted by the arrow in Fig.~\ref{fig:4}(c). We thus interpret the winged, asymmetric spectrum as a broad underlying distribution of relaxation timescales (well described by a power law at early times) compressed by dynamic facilitation at long times. Ironically, in our picture, the $\alpha$-peak itself is in ``excess'' of a much broader underlying time distribution with a high-frequency power law shape. In this view, the excess wing forms an integral part of the structural relaxation.

Our study frontally attacks a central question regarding the relaxation dynamics of supercooled liquids near the experimental glass transition and paves the way for many more studies of a totally unexplored territory now made accessible to modern computer studies. Enlarging further the family of available computer glass-formers would also help filling the gap with the more complex molecular systems studied experimentally.

\vspace{1cm}

\textit{Acknowledgments--}
 We thank G. Biroli, M. Ediger and J. Kurchan for discussions, and S. Nagel for detailed explanations about experiments. Some simulations were performed at MESO@LR-Platform at the University of Montpellier. This work was supported by a grant from the Simons Foundation (\#454933, L.B.), the European Research Council under the EU's Horizon 2020 Program, Grant No. 740269 (C.S.), a Herchel Smith Postdoctoral Research Fellowship (C.S.), a Ramon Jenkins Research Fellowship from Sidney Sussex College, Cambridge (C.S.) and Capital Fund Management -- Fondation pour la Recherche (B.G.).

\newpage

\section*{Methods}
\label{sec:methods}

\subsection*{Glass-forming computer models}

We study a non-additive, continuously polydisperse mixture of spherical particles of equal mass $m$ in two and three dimensions ($d=2,3$)~\cite{swap}. Two particles $i$ and $j$, at a distance $r_{ij}$ from one another interact via the repulsive potential 
\begin{equation}
v(r_{ij})=\epsilon\left(\frac{\sigma_{ij}}{r_{ij}}\right)^{12}+c_0 + c_2\left(\frac{r_{ij}}{\sigma_{ij}}\right)^2+ c_4\left(\frac{r_{ij}}{\sigma_{ij}}\right)^4,
\end{equation}
if $r_{ij}/\sigma_{ij} < x_c=1.25$. The constants $c_0=-28\epsilon/x_c^{12}, c_2=48\epsilon/x_c^{14}, c_4=-21\epsilon/x_c^{16}$ ensure continuity of the potential and its first two derivatives at the cutoff $x_c$.
The particles' diameters $\sigma_i$ are distributed from $\mathcal{P}(\sigma) =\mathcal{A}/\sigma^3$ with $\mathcal{A}$ a normalisation constant, $\sigma_\mathrm{max}/\sigma_\mathrm{min}=2.219$. We use the average diameter $\overline \sigma$ as unit length, $\epsilon$ as unit energy (the Boltzmann constant is set to unity) and $\sqrt{m\overline{\sigma}^2/\epsilon}$ as unit time. In these units, $\sigma_\mathrm{min} = 0.73$ and $\sigma_\mathrm{max} = 1.62$. We employ a non-additive cross-diameter rule $\sigma_{ij} =0.5(\sigma_i+\sigma_j)( 1-0.2|\sigma_i-\sigma_j|)$ to avoid fractionation and crystallization at low temperature~\cite{swap}. We simulate the glass-forming model at number density of particles $\rho=N/L^d = 1$ in a cubic/square box of linear size $L$ using periodic boundary conditions. We consider various system sizes: $N=1200, 10000$ in $3d$ and $N=2000, 10000$ in $2d$. 

\subsection*{Preparation of equilibrated configurations}

The model glass-forming liquid is efficiently simulated at equilibrium with the swap Monte Carlo algorithm. We employ the hybrid swap Monte Carlo/Molecular Dynamics algorithm implemented in the LAMMPS package ($2d/3d$) or homemade code ($3d$), with optimal parameters, as described in Ref.~\cite{lammpsswap}. We prepare $n_s \in \argc{200,450}$ independent equilibrated configurations at temperatures down to the experimental glass transition temperature.

\subsection*{Molecular dynamics simulations}

The equilibrium configurations generated by the swap algorithm are used as initial conditions for standard molecular dynamics (MD) simulations with integration time step equal to $0.01$. In $3d$, we run conventional MD (NVE) simulations and NVT simulations in $2d$ using a Nos\'e-Hoover thermostat. The simulations are either run using a homemade MD code or with the LAMMPS package, which allows us to run multi-CPU simulations and perform extremely long runs for relatively large systems ({\it e.g.}, two months on 24 CPUs for Fig.~2).

\subsection*{Relating experimental and numerical timescales}

We measure the relaxation time at the onset of glassy dynamics as reference time, and use this value to translate numerical timescales into experimental ones. In experiments, many supercooled liquids have $\tau_o \approx 10^{-10}$~s. We measure $\tau_o \approx 3$ in $2d$ and $3d$ simulations. In $3d$, the longest simulation time is $t_{\textrm{max}} = 1.5\times 10^{7} = 5\times 10^6 \tau_o$. We therefore simulate the equilibrium relaxation at $T_g$ over $0.5$~ms. In $2d$, we ran monthslong simulations to reach $t_{\textrm{max}}^{2d}  = 6\times 10^{8} = 2\times 10^{8} \tau_o$. Our numerical approach therefore allows us to observe the equilibrium dynamics over $20$~ms at $T_g$, which is a giant leap forward in equilibrium simulations of supercooled liquids.

\subsection*{Average dynamic observables}

In $3d$, we monitor the relaxation dynamics via the self-intermediate scattering function 
\begin{equation}
F_s(t) = \left\langle \frac{1}{N}\sum_{i=1}^N\cos \argc{ \bm{q}\cdot \delta \bm{r}_i(t) }\right\rangle_{\bm{q},n_s},
\end{equation}
where $\delta\bm{r}_i(t)$ is the displacement of particle $i$ over time $t$. The brackets indicate the ensemble average over $n_s$ independent runs along with an angular average over wavevectors with $|\bm{q}|=6.9$ (first peak in the total structure factor).

In $2d$, collective long-ranged fluctuations give rise to a spurious contribution to the displacements of particles~\cite{illing2017mermin} which affects the measurement of $F_s(t)$ and makes it ill-suited to capture the glassy slowdown. We instead study the dynamics through the evolution of the local environment of particles, instead of their displacements. We define a bond-orientational correlation function $C_{\Psi}(t)$~\cite{flenner2015fundamental}. We introduce the six-fold bond-orientational order parameter of particle $i$
\begin{equation}
\Psi _i (t) = \frac{1}{n_i} \sum_{j = 1}^{n_i} e^{\mathrm{i} 6\theta _{ij}(t)},
\end{equation}
where $n_i$ is the number of neighbours of $i$ at time $t$. Neighbours are particles $j$ with $r_{ij}<1.45$ (first minimum in the radial distribution function). Alternative definitions of neighbours, {\it e.g.}, via Voronoi tessellation or solid-angle based method~\cite{van2012parameter}, lead to the same quantitative results. Here $\theta _{ij}(t)$ is the angle between the $x$-axis and the axis connecting $i$ and $j$ at time $t$, without loss of generality thanks to rotational invariance. The bond-orientational correlation function is defined as 
\begin{equation}
C_{\Psi} (t) = \left\langle \frac{\sum_i \Psi _i (t) \argc {\Psi _i (0)}^* }{\sum_i | \Psi _i (0)|^2} \right\rangle_{n_s},
\end{equation}
where the brackets denote the ensemble average over $n_s$ independent runs, and the star is the conjugate complex. In $2d$, we define the relaxation time via $C_{\Psi}(\tau_{\alpha}) = e^{-1}$.

\subsection*{Mobility at the single-particle level}

When analysing the mobility at the single-particle level, we first need a criterion to distinguish between mobile and immobile particles. In $3d$, we have considered several mobility definitions which all give quantitatively similar results. The first mobility definition is based on displacements. To remove fast dynamical processes, we use the conjugate-gradient method and find the inherent structure (IS) of a configuration at time $t$, $\{\bm{r}_i^{IS}(t)\}$. Particle $i$ is defined as mobile at time $t$ if $|\bm{r}_i^{IS}(t) - \bm{r}_i^{IS}(0)| > 0.8$~\cite{schroder2000crossover}. This cutoff is between the first minimum and the second maximum of the self part of the van Hove function
$G_s(r,t)=\langle \delta(r-|\bm{r}_i^{IS}(t)|) \rangle_{i,n_s}$
in the time regime where $F_s(t)$ is almost constant. This first mobility definition is however not convenient in $2d$ because of the collective long-ranged fluctuations which affect the translational dynamics.

A second mobility definition is based on changes in the particle's local environment. At time $t=0$, we find the number $n_i$, and identity of particle $i$'s neighbours, defined as particles $j$ with $r_{ij}/\sigma_{ij}<1.485$ in $3d$ (1.3 in $2d$), corresponding to the first minimum in the rescaled pair correlation function $g(r_{ij}/\sigma_{ij})$. We define the bond-breaking correlation as the fraction of remaining neighbours at time $t$
\begin{equation}
C_B^i(t)=\frac{n_i(t|0)}{n_i},
\end{equation}
where $n_i(t|0)$ is the number of particles neighbour of $i$ at $t=0$ and still neighbour at $t$. To avoid short time oscillations in $C_B^i$ caused by particles frequently exiting/entering the shell defining neighbours, we use a slightly larger cutoff to define neighbours at $t>0$, namely $r_{ij}/\sigma_{ij}<1.7$ (in $d=2,3$). We compute the bond-breaking correlation function 
\begin{equation}
C_B(t) = \left\langle \frac{1}{N}\sum_{i=1}^N C_B^i(t) \right\rangle_{n_s},
\end{equation}
averaged over $n_s$ independent runs.

A particle is defined as mobile at $t$ if $C_B^i(t)<0.55$, \textit{i.e.}, if it has lost half of its initial neighbours. The cutoff value ensures that the set of particles identified as mobile in this way significantly overlap with that identified via the displacement criterion. We then introduce clusters of mobile particles. Two particles $i$ and $j$ mobile at time $t$ belong to the same cluster if $r_{ij} < 1.5$ in $3d$ and 1.4 in $2d$, close to the first minimum of $g(r)$. 

\subsection*{Relevant temperature scales}

\label{sec:temp_scale}

We determine three temperature scales relevant to the glassy slowdown: the onset temperature of glassy dynamics $T_o$, the mode-coupling crossover temperature $T_{\rm mct}$ below which conventional MD simulations cannot reach equilibrium, and the extrapolated experimental glass transition temperature $T_g$. In $3d$, $T_o=0.2, T_{\rm mct}=0.095, T_g=0.056$. In $2d$, $T_{o,2d}=0.2, T_{\rm mct,2d}=0.12, T_{g,2d}=0.07$. We fit the high-temperature $\ta$ data to an Arrhenius law, and identify the onset $T_o$ as the temperature below which $\ta$ is super-Arrhenius. We note $\tau_o = \ta(T_o)$. The mode-coupling crossover temperature $T_{\rm mct}$ is obtained by fitting the data with a power law $\tau_{\alpha}(T) \propto (T - T_{\rm mct})^{-\gamma}$ in the regime $0 \leq \log_{10}(\tau_{\alpha}/\tau_o) \leq 3$~\cite{gotze2008complex}, with $\gamma=2.7$ and $2.5$ in $d=2,3$ respectively. Given that $\log_{10} (\tau_{\alpha}/\tau_o) \approx  4$ at $T_{\rm mct}$, this temperature delimits the regime $ T > T_{\rm mct}$ where MD alone can reach equilibrium, from the regime $ T < T_{\rm mct}$ where the swap algorithm is needed to perform equilibrium simulations. The experimental glass transition temperature $T_g$ is defined by $\log_{10} (\tau_{\alpha}(T_g)/\tau_o) = 12$. In $3d$, the longest simulation time is $t_{\textrm{max}} = 1.5\times 10^{7} = 5\times 10^6 \tau_o$, so we can directly access $\log_{10} (\tau_{\alpha}/\tau_o) \lesssim 7$. We thus need to extrapolate our data over 5 decades to locate $T_g$. We increase the accuracy of the extrapolation by using time-temperature superposition (TTS), which is well-obeyed in our model~\cite{howtomeasure}. In the temperature regime where correlation functions reach $e^{-1}$, the second step of the relaxation is well-fitted by a stretched exponential $F_0e^{-(t/\ta)^\beta}$. The stretching exponent $\beta \simeq 0.56$ in $3d$ (in $2d$, $\beta\simeq 0.6$ for $C_{\Psi}$ and $\beta\simeq 0.67$ for $C_B$) is almost temperature-independent, and the amplitude $F_0$ slightly increases with decreasing temperature. Fixing $\beta$, we estimate $\ta$ at temperatures where decorrelation is sufficient to perform accurate TTS, extending our measurements over $\sim2$ decades. We extrapolate $\ta$ over the 4 remaining decades using an Arrhenius fit $\tau_\alpha(T)\propto e^{E_A/T}$ with $E_A = 2.67$ in $3d$ ($2.97$ in $2d$), and locate $T_g$. Importantly, the Arrhenius extrapolation is a safe choice as it at worst underestimates relaxation times. 

\subsection*{Computation of relaxation spectra}
\label{sec:spectra}

The computation of relaxation spectra $\chi''(\omega)$ first requires to differentiate the correlation function with respect to the logarithm of time. We use a first-order finite difference approximation. Namely, if configurations are stored at logarithmically-spaced times $\{t_k\}_{k=1\dots n}$, we have for $k>1$
\begin{equation}
\frac{\mathrm{d}F_s(t_k)}{\mathrm{d}\log t}=\frac{F_s(t_k)-F_s(t_{k-1})}{\log(t_k)-\log(t_{k-1})}.
\end{equation}
The integral in Eq.~(1) is then evaluated by
\begin{equation}
\chi''(\omega)=-\sum_{k=2}^n\frac{\mathrm{d}F_s(t_k)}{\mathrm{d}\log t}\frac{\omega t_k}{1+(\omega t_k)^2}\log\left(\frac{t_k}{t_{k-1}}\right).
\end{equation} 
We use the bond-breaking correlation function $C_B$ instead of $F_s$ in $2d$. In the Supplementary Information, we discuss errors which arise from computing the spectrum when $F_s$ does not decay to zero. We also discuss issues related to statistical noise and the comparison to direct Fourier transforms.

\subsection*{Trap model}

We consider traps with energy levels $E>0$ drawn from the exponential power distribution
\beq 
\rho(E) = \frac{\alpha}{E_0\Gamma(1/\alpha)}e^{-(E/E_0)^{\alpha}},
\label{eq:rho}
\eeq
and take $E_0=1$ in the following. We assume that dynamics at temperature $T$ is thermally activated. The energy $E$ of a trap is renewed after a Poisson-distributed timescale of mean $\langle \tau (E) \rangle = e^{ E/T}$. The equilibrium energy distribution at temperature $T$ is
\beq 
P_{\text{eq}}(T,E) = \frac{\rho(E) e^{ E/T}}{Z(T) }~~\text{where}~~ Z(T)  = \int_0^{\infty} \mathrm d  E \rho(E) e^{ E/T}.
\eeq

We monitor relaxation dynamics by computing the average persistence function $p(t)$. In the absence of dynamic facilitation, the persistence can be directly computed 
\beq 
p(t) = \int_{0}^{\infty} \mathrm d  E P_{\text{eq}}(T,E) e^{-t/ \langle \tau(E) \rangle}.
\label{eq:persistence}
\eeq

In the absence of dynamic facilitation, the average persistence is evaluated using Mathematica (NIntegrate, working precision 30). We then calculate the relaxation spectrum $\chi''(\omega)$ by following the procedure described previously, replacing $F_s(t)$ with the persistence $p(t)$. We compute the persistence $p(t)$ over a time interval large enough to observe full decorrelation, $[10^{-10},10^{70}]$ for $\alpha =1.1$, $T=0.629$, and minimise errors in the relaxation spectrum.

\subsection*{Simulations of the facilitated trap model}

We consider a system composed of $N$ traps. We initialise the simulation with an equilibrium condition by sampling the traps' energies directly from the equilibrium distribution $P_{\text{eq}}(T,E)$. Since the cumulative probability distribution of energies $\mathcal{C}_{\text{eq}}$ cannot be computed explicitly, we use Mathematica to evaluate it, and to numerically construct the reciprocal function $\mathcal{E}={{\mathcal C}_{\rm eq}}^{-1}$. For each of the $N$ traps, we generate $X$ uniformly distributed in $[0,1]$, and assign it an energy $E = \mathcal{E}(X)$. This procedure generates an initial condition in equilibrium. Each trap is assigned a renewal time exponentially distributed, with mean $e^{E/T}$. We initialise the persistence $p_i(t=0)$ of all traps to one.

The dynamics proceeds as follows. First, we identify the trap $i_o$ with the smallest renewal time $\tau_{\text{min}}$, which will relax first. We update all other traps by subtracting $\tau_{\text{min}}$ to their renewal time $\tau_i$. When the trap $i_o$ relaxes, its persistence is set to zero, $p_{i_o}=0$ and we give it a new energy value sampled from $\rho(E)$, and a new renewal time, as described above. 

This relaxation event then affects all other traps. We attempt to displace their energy by a random amount $\delta E$ (different for each trap) uniformly distributed in $\argc{- \frac{\Delta}{\sqrt{N}}, \frac{\Delta}{\sqrt{N}}}$: $E \rightarrow E' = E + \delta E$. The scaling with $N$ ensures that the resulting dynamics is independent on $N$. We then accept or reject this attempt in order to leave the equilibrium probability distribution $P_{\text{eq}}$ unchanged. To this end, we introduce an effective potential $V = - T \log P_{\text{eq}}$, and compute the change in effective potential $\delta V =  T(E'^{\alpha} - E^{\alpha}) - \delta E$. We then use the Metropolis filter: if $\delta V < 0$, the change in energy is accepted, otherwise, it is accepted with probability $
\exp(-\delta V/T)$. When accepted, we pick a new renewal time exponentially distributed with average $e^{E'/T}$. When the move is completed, we again determine which of the traps is the next one to relax, and proceed as before. 

We measure the average persistence $p(t) = \left\langle \sum_{i} p_i(t)/N \right\rangle$, where the brackets indicate average over independent runs, and where the sum runs over all traps. We simulate the dynamics of the model until the total persistence is equal to zero.

\bibliography{main.bib}

\end{document}


\title{Supplementary Information: Microscopic origin of excess wings in relaxation spectra of supercooled liquids}

\author{Benjamin Guiselin*}

\affiliation{Laboratoire Charles Coulomb (L2C), Universit\'e de Montpellier, CNRS, 34095 Montpellier, France}

\author{Camille Scalliet*}

\affiliation{Department of Applied Mathematics and Theoretical Physics, University of Cambridge, Wilberforce Road, Cambridge CB3 0WA, United Kingdom}

\author{Ludovic Berthier}

\affiliation{Laboratoire Charles Coulomb (L2C), Universit\'e de Montpellier, CNRS, 34095 Montpellier, France}

\affiliation{Yusuf Hamied Department of Chemistry, University of Cambridge, Lensfield Road, Cambridge CB2 1EW, United Kingdom}

\date{\today}

\maketitle

\setcounter{equation}{0}
\setcounter{figure}{0}
\setcounter{table}{0}
\setcounter{page}{1}
\renewcommand{\theequation}{S\arabic{equation}}
\renewcommand{\thefigure}{S\arabic{figure}}
\renewcommand{\bibnumfmt}[1]{[S#1]}
\renewcommand{\citenumfont}[1]{S#1}
\renewcommand\thesubsection{\arabic{subsection}}
\renewcommand\thesubsubsection{\arabic{subsection}.\arabic{subsubsection}  }

\section{Computation of relaxation spectra}
\label{sec:spectra}
%

\rev{
\subsection{Cutoff errors}
An accurate computation of $\chi''(\omega)$ using Eq. (9) of the Methods requires that  $\mathrm{d}C /\mathrm{d}\log t$ vanishes at the boundaries $t_1$ and $t_n$, set by the timestep $\mathrm{d}t$ and maximum simulation time $t_{\text{max}}$, respectively. Our short timestep and fine time spacing ensure this condition at short times. However, at low temperature, $C$ has not fully decorrelated after $t_{\text{max}}$. There is thus an error associated with cutting the computation of $\chi''$ via the time integral at $t_{\text{max}}$, where $C$ is still decreasing and $\mathrm{d}C /\mathrm{d}\log t \neq 0$. In order to minimise the cutoff errors, we compute $\chi''$ only in the range $\omega\in[2\pi/t_\mathrm{max},2\pi/\mathrm{d}t]$, at discrete angular frequencies $\omega_k=2\pi/t_k$ for $k=1\dots n$. Given that the computation of $\chi''(\omega)$ is dominated by times $t_k \sim 1/\omega$ [see Eq.~(9)], the error on the spectrum is negligible over most of the frequency range $\omega \gg \omega_\mathrm{min} = 2\pi/t_\mathrm{max}$. Only when $\omega < 10 ~\omega_\mathrm{min}$ (found empirically) do cutoff errors become noticeable, as evidenced by the slight bending downwards observed in Fig.~1(c) of the main text and Fig.~\ref{fig:2dspectra}. Indeed, the absent contribution of $\mathrm{d}C /\mathrm{d}\log t < 0$ at later times, combined with the negative sign in Eq.~(9) imply that the spectrum is underestimated close to the cutoff. This effect is visible, but only weakly affects the final few points at low frequencies in the spectra. This can be tested by using toy functions. Importantly, the cutoff errors underestimate the excess wing, so this effect does not alter our conclusions.}

\rev{
\subsection{Extrapolated $\alpha$-relaxation peaks} }
The extrapolated $\alpha$-relaxation peaks, \textit{i.e.}, the dashed lines in Fig.~1(c) of the main text, are also computed via Eq.~(9). We take the correlation function $C(t)$ to be the stretched exponential fit of $F_s(t)$ \rev{[or $C_B(t)$ in $2d$]}. To avoid cutoff errors in $\chi''$, we first evaluate $C(t)$ analytically on a very large time interval $[10^{-40},10^{40}]$.

\begin{figure}[t]
\begin{center}
\includegraphics[width=0.5\columnwidth]{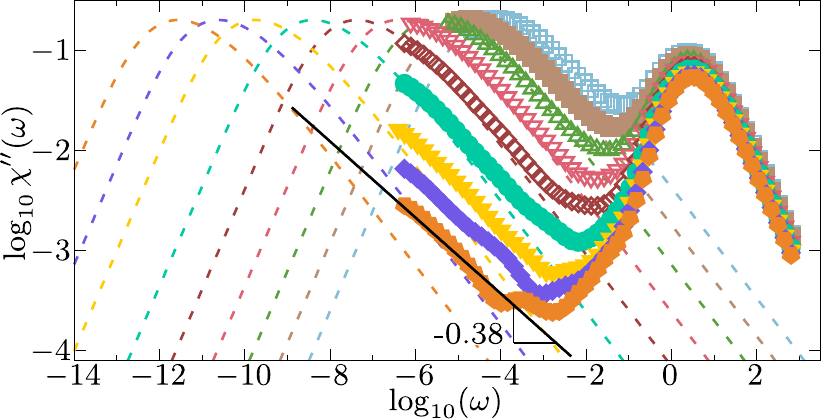}
\caption{\rev{\textbf{Raw relaxation spectra for the three-dimensional system.} In Fig.~1(c) of the main text, for the three lowest temperatures, we first applied a natural spline smoothing to $F_s$ in the time domain before computing $\chi''$. The ``bumps'' at low temperature/amplitude in this plot which disappear after the spline smoothing are due to statistical noise (despite extensive averaging), and have no physical origin.}}
\label{fig:wiggle}
\end{center}
\end{figure}

\rev{
\subsection{Smoothing of correlation functions at low temperature}
At the lowest temperatures, the correlation function $C$ decorrelates only very little over the course of the simulation, and appears almost constant [see Fig.~1(a) in the main text]. Despite our substantial computational effort to generate clean data (450 independent simulations at $T= 0.059$), there is still some statistical noise. The almost-plateau behaviour of $C$, combined with unavoidable noise in $C$ affect the evaluation of Eq. (9), and give rise to spurious oscillations in the resulting spectra. We report the spectra computed on the raw $F_s$ for the $3d$ system in Fig.~\ref{fig:wiggle}. We determined the typical error bar on the raw $F_s$ data in the time domain using the jackknife method over the independent initial samples. We then generated a Gaussian random time signal of zero mean and standard deviation given by this error bar, and computed the corresponding spectrum. The resulting spectrum was a randomly oscillating signal and we set the standard deviation of the latter as the typical error bar on our spectra, which turned out to be of order $4\times 10^{-4}$ [$\log_{10}(4\times 10^{-4})\approx -3.4$] and thus of similar amplitude as the one of the relative oscillations with respect to the underlying power law behaviour for the two lowest temperatures. Having demonstrated that the oscillations come from noise and bear no physical information, we decided to smooth the data to highlight the physically relevant features of the spectrum: the excess wing. For the three lowest temperatures in Fig.~1(c) in the main text, we applied a natural smoothing spline~\cite{spline} to $C$ after the ballistic regime in the time domain, from which we computed $\chi''$ in the Fourier domain. The resulting figure allows us to distinguish the physically relevant processes without being distracted by the oscillations stemming from the statistical noise. Indeed, different realisations of noise which is then added to the time correlation function do produce spectra with different oscillations but with similar features. In turn, the frequency at which the oscillations emerge is random and bear no physical origin.}

\rev{
\subsection{Alternative definition: direct Fourier transform}
We justify here our choice of procedure by comparing our results with that of a seemingly more natural (in the mathematical sense) choice: a direct Fourier transform 
\begin{equation}
\chi''(\omega)=-\int_0^{+\infty}\frac{\mathrm{d}C(t)}{\mathrm{d}t}\sin(\omega t)\mathrm{d}t.
\label{eq:fourier}
\end{equation}
To compute the above estimate of the spectrum for the $3d$ system, we first performed a cubic spline interpolation of our logarithmically-spaced data points for the self-intermediate scattering function $F_s$. We tested another interpolation scheme, the spline under tension, which yields similar results. We then evaluated the interpolated $F_s$ on a linear grid to compute the direct Fourier transform via Eq.~(\ref{eq:fourier}). This is necessary, as a direct Fourier transform of our logarithmically-spaced data yields extremely noisy spectra. The resulting spectra are presented in Fig.~\ref{fig:fourier}. As in Fig.~1(c) of the main text, the spectra at low temperature: i) are in excess of the $\alpha$-peak; ii) have an amplitude around 1\% of the $\alpha$-peak; iii) are well-fitted by a power law $\omega^{-0.38}$. Our main result thus does not depend on the specific procedure employed to compute $\chi'' (\omega)$. However, the direct Fourier transform is much costlier (several hours versus seconds of computation) and yields much noisier spectra compared to our procedure. The former is due to the large data sets (containing redundant information) involved to compute accurately the integral which contains a sine function. This comparative study demonstrates that the method that we employ efficiently yields quantitatively correct and smooth spectra, justifying our choice. 
\begin{figure}[t]
\centering
\includegraphics[width=0.5\textwidth]{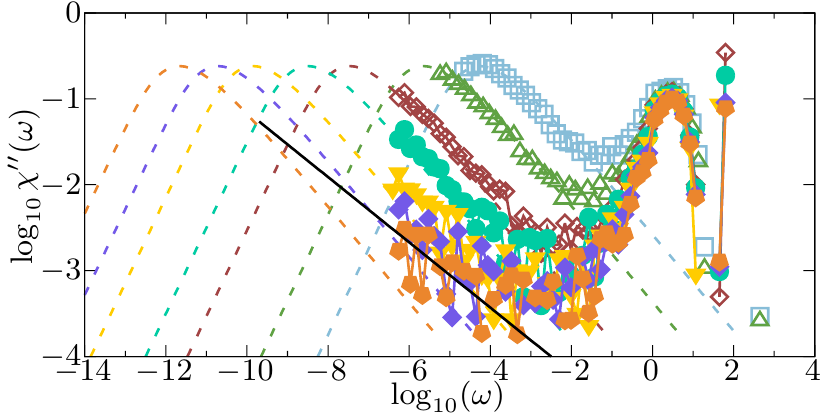}
\caption{\rev{\textbf{Spectra of the three-dimensional system computed via the direct sine transform.} These curves are computed from Eq.~(\ref{eq:fourier}). The color code is similar to Fig.~1(c) of the main text. The black line represents a power law $\chi''(\omega)\sim\omega^{-0.38}$.}}
\label{fig:fourier}
\end{figure}
}

%
%

\section{More results for the two-dimensional system}

\begin{figure}[b]
\begin{center}
\includegraphics[width=0.5\columnwidth]{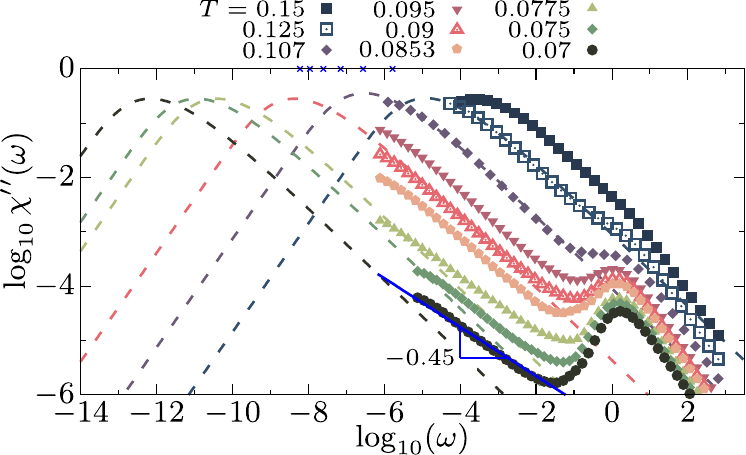}
\caption{\rev{\textbf{Relaxation spectra for the two-dimensional system.} Close to $T_{g,2d} = 0.07$, the spectra lie above the extrapolated $\alpha$-peak (dashed), and display a power law behaviour $\chi'' \sim \omega^{- \sigma_{2d}}$ with an exponent $\sigma_{2d} \approx 0.45$ (blue full line). The similarity between these $2d$ spectra and those shown for $d=3$ in Fig.~1(c) of the main text is clear. The crosses on the top axis indicate $1/t_{\text{snap}}$ for the snapshot times of Fig.~2 in the main text.}}
\label{fig:2dspectra}
\end{center}
\end{figure} 

\begin{figure}[t]
\begin{center}
\includegraphics[width=0.5\columnwidth]{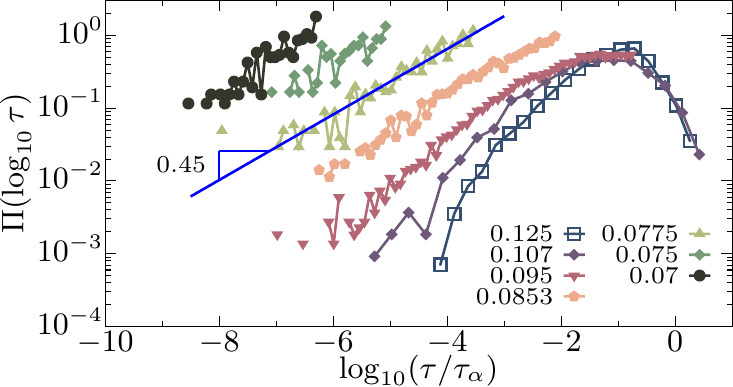}
\caption{\rev{\textbf{Waiting-time distribution of newly relaxing clusters in the two-dimensional system.} The color code is the same as in Fig.~\ref{fig:2dspectra} and the legend indicates the temperature $T_{2d}$. Approaching $T_{g,2d}$, the distribution develops a power law behaviour at $\tau \ll \ta$, with an exponent $0.45$ matching that of the excess wing in the spectra in Fig.~\ref{fig:2dspectra}. Here again, one recognizes the same behaviour as the three-dimensional results presented in Fig.~3 of the main text.}}
\label{fig:2ddistrib}
\end{center}
\end{figure}

We present in this section the results obtained for the two-dimensional model. We show in Fig.~\ref{fig:2dspectra} the relaxation spectra and in Fig.~\ref{fig:2ddistrib} the waiting-time distribution measured in the $2d$ model. They parallel the Fig.~1(c) and Fig.~3 presented in the main text for the $3d$ model.

We computed the relaxation spectra $\chi''$ from the bond-breaking correlation function $C_B$ defined in the Methods section of the main text. We measured the relaxation spectra using samples of $N=2000$ particles, averaged over $n_s=300$ independent runs, for most temperatures. At the lowest temperature $T_{2d} = 0.07 \simeq T_{g,2d}$, we needed very large statistics, $n_s=40$ simulations with $N=64000$ particles. The resulting spectra are shown in Fig.~\ref{fig:2dspectra} from $T_{2d} = 0.15$, above the mode-coupling crossover temperature ($T_{{\rm mct},2d} = 0.12$), down to the extrapolated glass transition temperature $T_{g,2d} = 0.07$. For some temperatures, we indicate with dashed lines the extrapolated $\alpha$-relaxation peak, following the same procedure as for the $3d$ model. Here, the stretching exponent is $\beta_{2d} = 0.67$. We find that the $\alpha$-peak coincides with the measured data at high temperature, but starts to deviate from $T_{2d}=0.0775$. At low temperature, the signal lies above the extrapolated $\alpha$-peak, and is well-fitted by a power law $\omega^{-\sigma_{2d}}$, with $\sigma_{2d}=0.45$. While some details of the spectra are different from the $3d$ model, an excess wing is also clearly observed in $2d$ simulations. A notable difference with $3d$ is the absence of a systematic high-frequency microscopic peak as the bond-breaking correlation does not display a plateau at intermediate timescales at large temperatures. However, this does not affect the excess wing.

The figure presents in particular the spectrum at $T_{2d}=0.09$, corresponding to the snapshots of Fig.~2 of the main text. At the top of Fig.~\ref{fig:2dspectra} we indicate with blue crosses the inverse time of the snapshots. While the time/frequency correspondence has its caveats, this shows to which stage of relaxation corresponds a given snapshot. Since producing the snapshots required monthslong simulations, we performed only one run and could not compute the spectrum accurately on a similarly broad frequency range, which explains why the symbols are located at lower frequencies than the spectrum. At this temperature $T_{2d}=0.09$, the $\alpha$-peak still interferes with the measured data, and an excess wing is not clearly observed. While little relaxation has taken place at the first snapshot of Fig.~2 of the main text, the corresponding signal is comparatively quite high in the spectrum. This demonstrates that the excess wing, observed at lower temperature, corresponds to the very beginning of the relaxation process.

We also measured the waiting time distribution for the appearance of new relaxed clusters in $2d$, and present the results in Fig.~\ref{fig:2ddistrib}. We show the data from temperatures close to the mode-coupling crossover temperature, down to the glass transition temperature. The temperature evolution of the distribution closely follows that of the $3d$ model. Below $T_{2d} = 0.095$ we cannot measure the entire distribution, which is thus determined up to an uninteresting prefactor. Close to $T_{{\rm mct},2d} = 0.12$, the distribution is already broad, extending down to $10^{-4}\ta$, and peaks around one decade before the $\alpha$-relaxation time. As the temperature decreases, the distribution broadens towards shorter and shorter times compared to the $\alpha$-relaxation time, indicating that relaxation starts much before the bulk. For temperatures $T_{2d}<0.0853$, the distribution exhibits a clear power law behaviour over more than four decades in time at $\tau \ll \ta$. The distribution is well-fitted by a power law $\Pi(\log_{10}\tau)\sim\tau^{0.45}$, which is the same exponent observed for the excess wing in the relaxation spectra. The matching of both exponents indicates that the excess wing observed in the spectra originates from very early relaxation events which take place over power law distributed timescales. Our main conclusions drawn in the main text for the $3d$ results therefore hold also for the $d=2$ model. This suggests a universal explanation of excess wings in terms of very early relaxation events occurring over broadly distributed timescales.

\bibliography{si.bib}